\documentclass[conference]{IEEEtran}
\IEEEoverridecommandlockouts
\usepackage{amsmath,amssymb,amsfonts}
\usepackage{algorithmic}
\usepackage{graphicx}
\usepackage{textcomp}
\usepackage{xcolor}
\usepackage{siunitx}
\usepackage[capitalise]{cleveref}
\usepackage{threeparttable}
\usepackage{url}

\ifCLASSOPTIONcompsoc
\usepackage[caption=false,font=normalsize,labelfon
t=sf,textfont=sf]{subfig}
\else
\usepackage[caption=false,font=footnotesize]{subfi
g}
\fi

\usepackage{myTikz}

\usetikzlibrary{external}
\def\BibTeX{{\rm B\kern-.05em{\sc i\kern-.025em b}\kern-.08em
    T\kern-.1667em\lower.7ex\hbox{E}\kern-.125emX}}
    
\begin{document}

\title{TCAD Simulations of Radiation Damage in 4H-SiC%
\thanks{Jürgen Burin has been supported by the Austrian Research Promotion Agency (FFG) in the project RadHardDetSim (OFFG000345).}
}

\author{%
\IEEEauthorblockN{Jürgen Burin, Christopher Hahn, Philipp Gaggl, Andreas Gsponer, Simon Waid and Thomas Bergauer}\\
\IEEEauthorblockA{\textit{Institute of High Energy Physics} \\
\textit{Austrian Academy of Sciences}\\
1050 Vienna, Austria \\
juergen.burin@oeaw.ac.at}
}

\maketitle

\begin{abstract}
To increase the scientific output of particle physics experiments, upgrades are underway at all major accelerator facilities to significantly improve the luminosity. Consequently, the solid-state detectors used in the experiments will exhibit more severe radiation-induced damage. To ensure sufficiently long sensor lifetimes, alternative materials to the established silicon sensors, with improved resilience to radiation, are investigated. For one of the promising candidate materials, silicon carbide, only recently a model describing the radiation damage in technology aided computer design (TCAD) simulations has been proposed.

In this paper we present our latest achievements towards modeling radiation damage of 4H-SiC in TCAD tools. We first verify the utilized TCAD framework against published silicon data and then use it to approximate measurements of neutron-irradiated 4H-SiC particle detectors. We are able to confirm in simulations the measurement results, i.e., an almost flat capacitance as a function of bias voltage and a decreasing forward current with increasing particle fluence. Based on our simulations we are able to explain the latter by trapped charge carriers that create a space charge region within the device.
\end{abstract}

\begin{IEEEkeywords}
TCAD simulations, silicon carbide (SiC), particle detector, radiation damage
\end{IEEEkeywords}

\section{Introduction}
\label{sec:introduction}

Particle accelerators, for example those at the research center CERN in Geneva, are a crucial component to explore the limits of the standard model of physics. In order to develop proper extensions, scientists use information that are gathered from very rare particle collision events. To improve the scientific output of the collider experiments at CERN the High-Luminosity upgrade for the LHC (HL-LHC) is currently in preparation and will be deployed in 2029. As the name indicates the purpose of this enhancement is an increase in luminosity, which, however, also leads to more radiation damage in the detectors. In order to avoid frequent replacement of detectors and a performance loss during operation, materials that feature an improved radiation hardness and can serve as a basis for future particle detectors are highly desired. One very promising candidate is silicon carbide (SiC). Due to its material properties, such as a wide bandgap (low leakage current) and high charge carrier mobilities (fast devices), devices outperforming the existing silicon ones are enabled.

The key question in using SiC, or more specifically, the most commonly used polytype 4H-SiC, for HEP experiments is in which degree the performance degrades while being exposed to ionizing radiation. Recent measurements~\cite{Gsponer2023} revealed that at fluences higher than \SI{e15}{n_{eq}/\cm^2} the capacitance of a diode is absolutely constant from highly positive to highly negative bias (see~\cref{fig:CV_measurements}). Furthermore, it is possible to heavily forward bias the device without seeing significant conduction (see \cref{fig:IV_measurements}). In order to understand these observations, technology computer aided design (TCAD) simulations can be beneficial. Modeling the internal structure of the particle detectors, i.e., layer thicknesses and doping profiles, enables a time- and cost-effective evaluation of possible candidates of radiation damage, such as traps and recombination centers.

In this paper, we will thus present our current efforts toward modeling radiation damage in 4H-SiC by means of TCAD simulations. After verifying the feasibility of our simulation approach by comparing the results to existing data for silicon sensors, we focus on recreating measurements of 4H-SiC. Qualitatively, good matches are achieved and the simulation results allow us to hypothesize possible causes for the observed phenomena. For a quantitative fit, however, further improvements are required. In this publication, we first provide some background information in \cref{sec:background} before showing the achieved simulation results in \cref{sec:simulations} and concluding the paper in \cref{sec:conclusion}.

\begin{figure*}[t]
    \centering
    \subfloat[capacitance-voltage]{
         \resizebox{0.45\linewidth}{!}{%
            \includegraphics{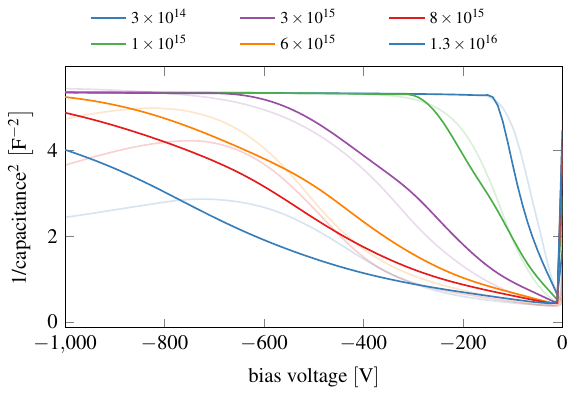}
            }
         \label{fig:Si_CV}
    }
    \hfil
    \subfloat[electric fields from anode (\SI{0}{\um}) to cathode (\SI{200}{\um}) at a reverse bias of \SI{1}{\kV}]{
         \resizebox{0.49\linewidth}{!}{%
            \includegraphics{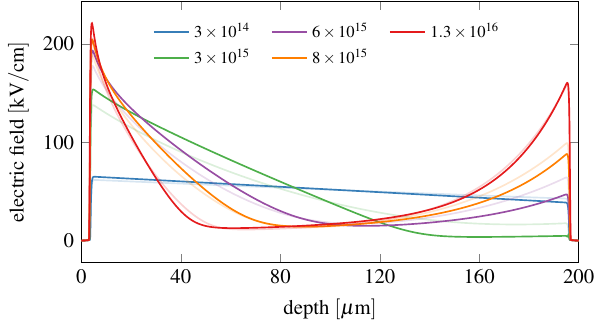}
            }
         \label{fig:Si_fields}
    }
    \caption{Simulations of a silicon diode for fluences from \SIrange{3e14}{1.3e16}{n_{eq}/\cm^2}. Solid lines represent the simulation results from GTS and opaque lines those from Sentaurus. For a better fit, the fluences in the GTS simulations of the capacitance have been scaled with a factor of \num{0.3}.}
\end{figure*}

\section{Background}
\label{sec:background}

Silicon carbide (SiC) is a wide bandgap material that is heavily deployed in power electronics. It can crystallize in different crystal arrangements, called polytypes, whereas 4H is the most commonly used in industry (bandgap energy $E_g = \SI{3.285}{\eV}$~\cite{freitas1995photoluminescence}). Due to its high charge carrier mobility and thermal conductivity, it is a very good alternative to the currently established silicon.

The most basic particle detectors are reverse biased p-i-n diodes. Charge carrier pairs generated by the impinging particle in the space charge region are, due to the high electric field in this region, immediately drawn towards the cathode (electrons) resp. anode (holes) and induce a current signal during their drift. For detection purposes the wide bandgap of 4H-SiC is not solely advantageous as fewer charge carriers are generated per unit distance. In the case of a minimum ionizing particle (MIP), the overall amount is less than $100$ electron-hole pairs per micron~\cite{Christanell_2022}. Increasing the device thickness and, therefore, the active volume would strengthen the signal. However, the achievable layer thicknesses are still constrained by the limitations in growing high-quality epitaxial layers. Consequently, sophisticated electronics is required to prepare the signals before post-processing.

Further constraints on the device thickness are given by the high intrinsic doping of approximately \SI{e14}{\cm^{-3}} achieved in industry, which increases the full depletion and operational voltage drastically compared to silicon. For currently available technologies approximately \SI{500}{\V} reverse bias is required to fully deplete a \SI{50}{um} thick device. Even though 4H-SiC has a very high breakdown field strength, such high voltage values imply further design challenges such as the need for guard structures.

To characterize the behavior of particle detectors, three measurements are commonly conducted in high energy physics: I-V, C-V, and charge collection efficiency (CCE). Current measurements (I-V) in backward bias extract the dark current, which limits the detection sensitivity. In 4H-SiC the thermal charge carrier generation is, however, so minimal that the current is below \SI{1}{\pA} and limited experimentally by leakage paths and the resolution of the measurement device. The current in forward bias helps to classify the radiation damage, as described in the next section. Capacitance measurements (C-V) are used to evaluate the space charge region (active volume of the device) which can be used to extract the electrically active doping. Finally, the CCE is used to describe the ratio of a collected signal to the one achieved in a fully depleted and defect free device. Since the integral over the transient current induced by the moving charge carrier is equal to the deposited charge, the CCE denotes how efficient charge carriers can be collected.

\begin{figure*}[t]
    \centering
    \subfloat[measurements~\cite{Gsponer2023}]{
        \centering
         \resizebox{0.47\linewidth}{!}{%
            \includegraphics{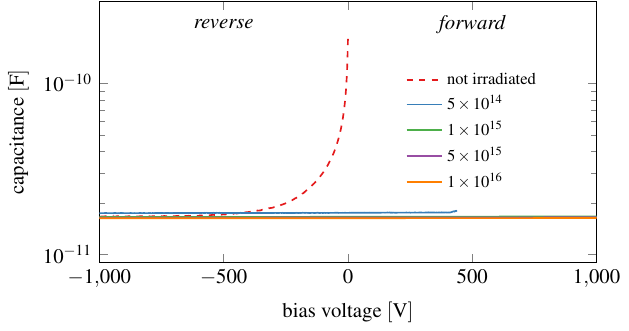}
            }
         \label{fig:CV_measurements}
    }
    \hfil
    \subfloat[GTS simulations]{
         \centering
         \resizebox{0.47\linewidth}{!}{%
            \includegraphics{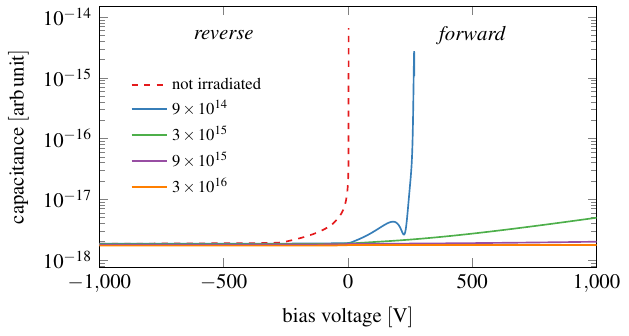}
            }
         \label{fig:4H-SiC_CV}
    }
    \caption{Capacitance-voltage (C-V) characteristics of a 4H-SiC diode for fluences from \SIrange{5e14}{3e16}{n_{eq}/\cm^2}. In the simulations lower fluences resulted in nonphysical results and had to be discarded.}
\end{figure*}

To model the radiation damage in TCAD simulations, I-V, C-V, and CCE measurements of well known devices represent a crucial input. For this purpose we used the results reported by Gsponer \textit{et al.}~\cite{Gsponer2023} who measured 4H-SiC detectors irradiated with \SI{1}{\MeV} equivalent neutron fluences between \SI{e14}{n_{eq}/\cm^2} and \SI{e16}{n_{eq}/\cm^2}. The measurements revealed the surprising fact that the capacitance, which should change due to the varying width of the space charge region, stays constant in forward and reverse bias conditions (cp. \cref{fig:CV_measurements}). In addition, the diodes can be heavily biased in forward direction without showing significant conductance. While the onset of forward conductance in regular diodes starts around \SI{2}{\V} the radiated ones can even withstand more than \SI{1}{\kV} (cp. \cref{fig:IV_measurements}). 

In literature the radiation damage in 4H-SiC has been investigated in several publications, however, the achieved results vary significantly. Overall, carbon vacancies, i.e., missing carbon atoms in the lattice, are the cause of the most prominent charge carrier traps Z\textsubscript{1,2} and EH\textsubscript{6,7}. While the former is close to the conduction band~\cite{klein2009}, the latter is located in the middle of the bandgap~\cite{megherbi2018a,zhang2003}. Besides the energy level, the type of the trap (donor or acceptor) and the cross sections for electrons and holes, i.e., the probability of a respective charge carrier entering the trap, are of importance. Unfortunately, the corresponding values in the literature still show a considerable spread, making a proper selection very challenging.

\section{Simulations}
\label{sec:simulations}

In this section, we present the TCAD simulation results, which have been achieved using the simulation framework from Global TCAD Solutions (GTS)~\cite{GTS} in version 2024.03 Build 25. To verify the results we first recreated the radiation damage simulations by Schwandt \textit{et al.}~\cite{schwandt2019} for silicon. Afterwards, we turned to modeling the radiation damage in 4H-SiC. Despite claiming that I-V, C-V and CCE curves are used to characterize a particle detector in \cref{sec:background} we will not explicitly show the achieved CCE results here due to the tight page limit.

During our simulations with the GTS framework we encountered multiple convergence issues, which can be traced back to the still ongoing optimization of the tools. Consequently, we had to slightly adapt the simulation setup for each of the investigated curves. In the case of 4H-SiC, for example, incomplete ionization had to be disabled and the simulation grid was fine-tuned for forward and reverse bias separately.

\subsection{Verification of Silicon}

For the simulations of silicon, we used the Hamburg Penta Trap Model (HPTM) proposed in~\cite{schwandt2019}. We want to highlight that we slightly simplified the structure of the investigated device, especially the doping profiles, so our results are not directly comparable to~\cite{schwandt2019}. In detail a \SI{200}{\um} thick 1D model of a pin-diode (bulk doping \SI{3.0e12}{\cm^{-3}}) with Gaussian-distributed doping profiles for the n+ ($N_D = \SI{1.0e19}{\cm^{-3}}, \sigma = \SI{1}{\um}$) and p+ ($N_A = \SI{1.0e19}{\cm^{-3}}, \sigma = \SI{1}{\um}$) contacts was used.

The simulation results of GTS fit qualitatively well to results achieved by simulations in Synopsys\textsuperscript{\textregistered} Sentaurus~\cite{Sentaurus}. However, also differences, especially in regard to the radiation dose, were observed. For example, almost identical capacitance values have been achieved but for different fluences. In fact, scaling the latter by a factor of \num{0.3} led to a very good match (see \cref{fig:Si_CV}). In some case GTS even outperforms Sentaurus. For high reverse bias and neutron fluences above \SI{3e15}{n_{eq}/\cm^2} the latter predicts an increase of the device's capacitance, which does not match the measurement results shown in~\cite{schwandt2019}.

A key advantage of simulations compared to measurements is the ability to look inside the device. In \cref{fig:Si_fields} we show the simulated electric field across a reverse biased diode. Clearly visible is a double peak structure, whereat the field in the middle is almost zero. This can be explained by bulk traps that bind the majority charge carriers (electrons close to the cathode and holes close to the anode) and thus create a (quasi-stationary) space charge. 

Although no perfect match among the tools was achieved, we conclude that the GTS framework is viable to provide qualitative trends in radiation damage. The quantitative differences can be, partially, explained by deviating implementations of the respective algorithms in each tool.

\subsection{4H-SiC Radiation Simulations}

\begin{figure*}[t]
    \centering
    \subfloat[measurements~\cite{Gsponer2023}]{
         \centering
         \resizebox{0.45\linewidth}{!}{%
            \includegraphics{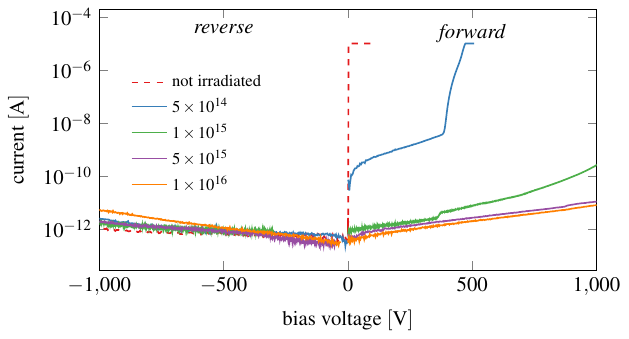}
            }
         \label{fig:IV_measurements}
    }
    \hfil
    \subfloat[GTS simulations]{
         \centering
         \resizebox{0.45\linewidth}{!}{%
            \includegraphics{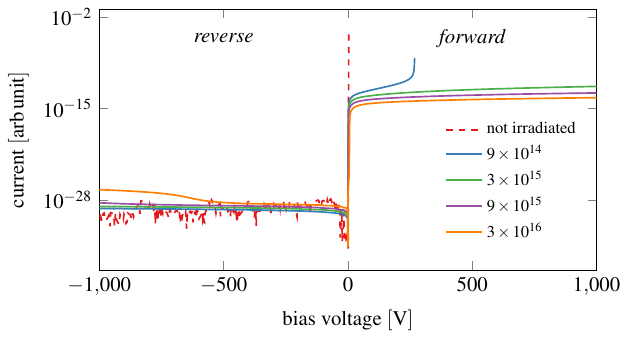}
            }
         \label{fig:4H-SiC_IV}
    }  
    \caption{Current-voltage (I-V) characteristics of a 4H-SiC diode for fluences from \SIrange{5e14}{3e16}{n_{eq}/\cm^2}. In the simulations lower fluences resulted in nonphysical results and had to be discarded.}
\end{figure*}

After the successful verification against silicon, our next goal is to approximate the measurements shown in \cref{fig:CV_measurements} and \cref{fig:IV_measurements}. For this purpose we use the parameters shown in \cref{tab:4HSiC_parameters}, which is a simplified version of the model proposed in~\cite{Gaggl2024_Pisa}.

\begin{table}[b]
    \sisetup{output-exponent-marker = \text{e}}
    \begin{threeparttable}
        \centering
        \setlength{\tabcolsep}{5pt}
        \renewcommand{\arraystretch}{1.1}
        \caption{4H-SiC radiation damage model used in the simulations. $E_C$ and $E_V$ denote the conduction resp. valence band energy, $g_{int}$ the introduction rate and $\sigma_{e,h}$ the electron resp. hole cross-section.}
        \begin{tabular}{*{6}{c}}
             Defect & Type & Energy & $g_{int}$ & $\sigma_e$ & $\sigma_h$ \\
             &&& [\si{\per\cm}] & [\si{\cm^2}] & [\si{\cm^2}] \\ \hline \\[-7pt]
             Z\textsubscript{1,2} & Acceptor & $E_C - \SI{0.67}{\eV}$\,\tnote{a} & \num{5.0}\,\tnote{b} & \num{2e-14}\,\tnote{a} & \num{3.5e-14}\,\tnote{a} \\
             EH\textsubscript{6,7} & Donor\,\tnote{c} & $E_C - \SI{1.6}{\eV}$\,\tnote{d,e} & \num{1.6}\,\tnote{b} & \num{9e-12}\,\tnote{e} & \num{3.8e-14}\,\tnote{d,e} \\
             EH\textsubscript{4} & Acceptor & $E_C - \SI{1.03}{\eV}$\,\tnote{f,g} & \num{2.4}\,\tnote{b} & \num{5e-13}\,\tnote{g} & \num{5.0e-14}\,\tnote{g} \\[4pt] \hline
        \end{tabular}
        \label{tab:4HSiC_parameters}
        \begin{tablenotes}[para]
            \footnotesize
            \item[a] \cite{klein2009}
            \item[b] \cite{hazdra2014}
            \item[c] \cite{hornos2011}
            \item[d] \cite{megherbi2018a}
            \item[e] \cite{zhang2003}
            \item[f] \cite{hemmingsson1997}
            \item[g] \cite{alfieri2005}
        \end{tablenotes}
    \end{threeparttable}      
\end{table}

The C-V and I-V simulation results are shown in \cref{fig:4H-SiC_CV} and \cref{fig:4H-SiC_IV}, respectively. For the former the trend from the measurements can be qualitatively recreated. The decrease of the capacitance for increasing reverse bias vanishes completely. Instead, it stays at a constant low value, which only slightly increases at large current densities in forward bias. For fluences below \SI{9e14}{n_{eq}/\cm^2}, the simulation fails to deliver reasonable results as the capacitance even becomes negative.

\begin{figure}[t!]
    \centering
    \resizebox{1\linewidth}{!}{%
    \includegraphics{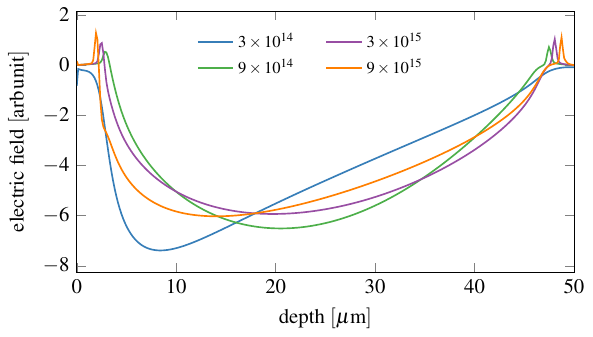}
    }
    \caption{Electric field along the 4H-SiC diode from anode (\SI{0}{\um}) to cathode (\SI{50}{\um}) in GTS for a forward bias of \SI{200}{\V} and fluences from \SIrange{3e14}{9e15}{n_{eq}/\cm^2}. The traps bind majority charge carriers and thus forms a space charge that prevents further charge carrier injection.}
    \label{fig:4HSiC_fields}
\end{figure}

The I-V simulation results in \cref{fig:4H-SiC_IV} show a difference of more than ten orders of magnitude between forward and reverse bias. Since the former is already in the \si{\pA}-range for measurements, one can envision that it is impossible to determine true dark current of the diode in the lab experimentally, due to edge/surface effects and current leakage paths. In the forward direction, the significant decrease of conductance is clearly visible. Compared to the reverse bias simulations shown earlier, the electric field in forward bias (see \cref{fig:4HSiC_fields}) paints a very different picture. Although, again, field peaks at both sides of the pin-diode are visible, the field drops in between below zero. Our explanation is that, as before, majority charge carriers get trapped near their respective contact, forming a space charge region. The lower the trap occupation the lower the lifetime of the electrons and holes. For sufficient forward bias, enough carriers are injected to fill the traps, which allows the former to move further into the device, significantly increasing the recombination current.

\section{Conclusion}
\label{sec:conclusion}

In this paper we presented our ongoing efforts towards modeling radiation damage of 4H silicon carbide (SiC) in technology computer aided design (TCAD) simulations. We first verified the utilized simulation framework from Global TCAD Solutions (GTS) against published results for silicon before turning to 4H-SiC. The achieved data fit existing ones qualitatively well, showing a decreasing device performance with higher particle fluences. Based on simulation results, e.g., the electric field deep inside the devices, we were able to retrace the lack of conductance of diodes in forward bias to trapped majority charge carriers.

Future research will be focused on a more thorough analysis of the introduced trap parameters, i.e., energy level, type, and cross-section, as literature values show large deviations. We also plan to investigate the trapping mechanisms in forward bias to a greater extend and verify our explanations by further measurement methods, such as DLTS, on additional and complementary devices.

\bibliographystyle{IEEEtran}
\bibliography{IEEEabrv,literature}

\end{document}